# MPTL'20 Proceedings:
# Creating Electronic Books-Chapters for Computers and Tablets Using Easy Java/JavaScript Simulations, EjsS Modeling Tool

*Loo Kang Wee[1]*
*Lawrence_Wee@moe.gov.sg*
*[1]Ministry of Education, Singapore*

*Abstract*
This paper documents the process (tools used, design principles derived and modeling pedagogy implemented) of creating electronic book-chapters (epub3 format) for computers and tablets using Easy Java/JavaScript Simulations (formerly EJS, new EjsS) and Modeling Tool. The theory underpinning this work is that learning by doing through dynamic and interactive simulation-models is more easily integrated than knowledge gained through static printed materials.
I started by combining related computer models with supporting texts and illustrations into a coherent chapter. From there, a logical next step towards more complete support for teachers and students is developing prototypes for electronic chapters on the topics of Simple Harmonic Motion and Gravity. This also is customized for the Singapore-Cambridge General Certificate of Education Advanced Level (A-level). I aim to inspire more educators to create interactive and open educational resources[1] for the benefit of all.
Prototypes:
http://iwant2study.org/ospsg/index.php/interactive-resources/physics/02-newtonian-mechanics/08-gravity/154-e-book-shm
http://iwant2study.org/ospsg/index.php/interactive-resources/physics/02-newtonian-mechanics/09-oscillations/153-e-book-gravity
Also available on Apple iBook Store, Google Play Book and Kindle Book.

## 1. Introduction

After the Singapore Easy Java/JavaScript simulations and EjsS Modeling Tool workshop[2] organised by Francisco Esquembre and Wolfgang Christian on November 25-28 2014, I started combining related computer models with supporting texts and illustrations into a coherent chapter. a logical next step towards tighter support for teachers and students. This paper aims to articulate some of the design ideas and tools used, as well as a mathematical modeling approach (Wee, 2014) developed to work in EjsS, similar to Tracker's video analysis kinematics model (Brown, 2012; Wee, Chew, Goh, Tan, & Lee, 2012; Wee & Leong, 2015; Wee, Tan, Leong, & Tan, 2015). Some examples in the e-chapter will be mentioned and, in conclusion, I hope to inspire more educators to create interactive and open educational resources (ISKME, 2008) for the benefit of all.

## 2. Tools Used

|   | **Computer Tools** | **Purpose** |
|---|---|---|
| **1** | EjsS Modeling Tool | Make interactive models, generate epub |
| **2** | BlueGriffon xhtml editor | Create remove markuperrors *.xhtml files with texts, images (recommend 1024x768), insert equations MathML |
| **3** | Mathtowebonline MathML equation editor | Generate equations in MathML format |

*Table. 1. Tools used and their respective purposes towards creating interactive textbook chapters.*

---

[1] http://www.unesco.org/new/en/communication-and-information/access-to-knowledge/open-educational-resources/what-are-open-educational-resources-oers/

[2] https://girep.org/newsletters/newsletter_2015_05.pdf page 10

Table 1 shows the tools used to create these interactive epub3 format electronic text book chapters, where EjsS Modeling Tool (Esquembre, 2012) is heavily relied upon to generate these *.xhtml format texts and JavaScript simulations. Since EjsS Modeling Tool does not create the *.xhtml format texts, but rather simply packages it with the JavaScript simulations, a separate xhtml editor such as BlueGriffon is used to create these texts with markup errors removed. Moreover, to generate free MathML format equations, I recommend using MathtoWebonline to create these equations and paste them into the *.xhtml files-texts-equations inside the BlueGriffon editor. Authors who wish to publish their e-chapters on Apple iBook should note that pictures need to size exactly to 1024x768 in dimensions.

### 3. Simulation Design Principles for electronic book-chapters

In the process of creating a suite of simulations for 2 chapters on Simple Harmonic Motion and Gravity (Wee & Goh, 2013) customized for the Singapore-Cambridge General Certificate of Education Advanced Level (A-level), three design principles (Wee, 2012; Wee & Goh, 2013; Wee, Lee, Chew, Wong, & Tan, 2015; Wee & Ning, 2014) emerged while preparing this paper; they are listed below.

### 3.1. User design: Simple & Optimum View of Screen

To create a user interface and view for optimized intellectual performance by students, cognitive load theory[3] (Roth, 1999) suggests the simulations to be designed with a simple layout with less user control interface. Figure 1 below shows a simple 2 panel layout and a bottom control panel that was well received by teachers and students in Singapore.

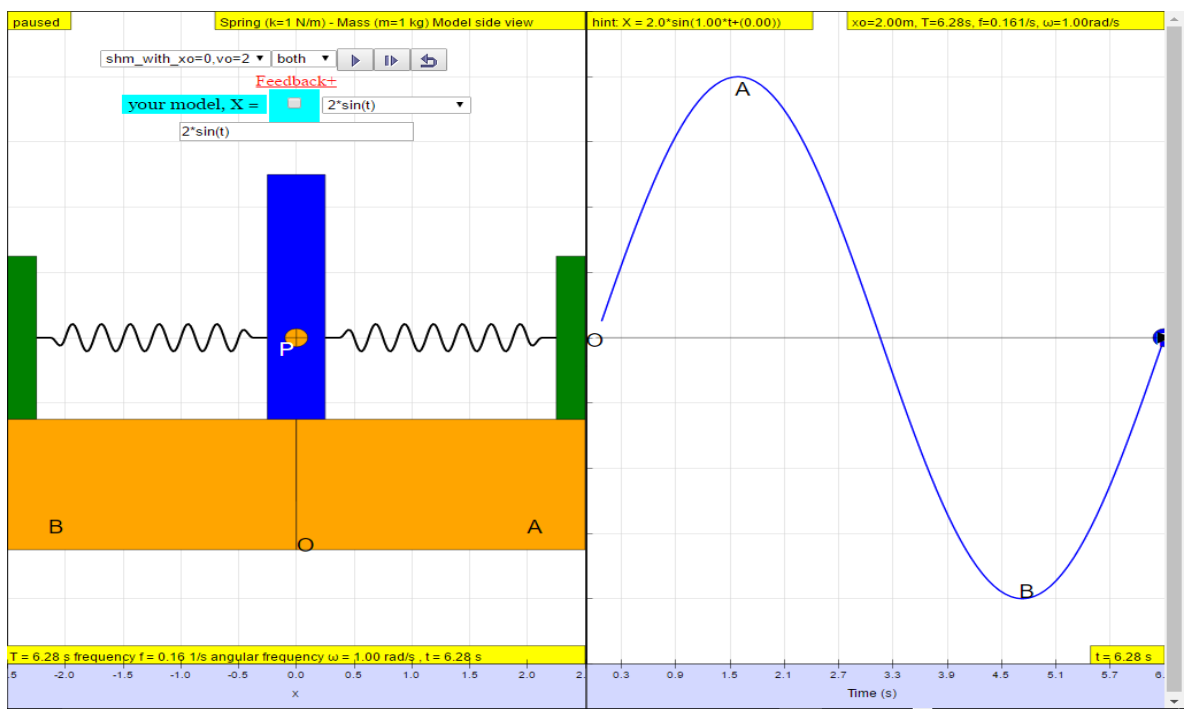

*Figure 1: Horizontal Spring Mass Model [4] with a simple world view layout on the left with only one other scientific view of displacement, x versus time, t graph.*

---

[3] https://en.wikipedia.org/wiki/Cognitive_load
[4] http://iwant2study.org/ospsg/index.php/interactive-resources/physics/02-newtonian-mechanics/09-oscillations/69-shm04

In addition being simple in view, I have implemented full screen capability on Android to allow tapping on the smaller screen size of students' personal hand phones. These devices can then be used for learning anytime and anywhere.

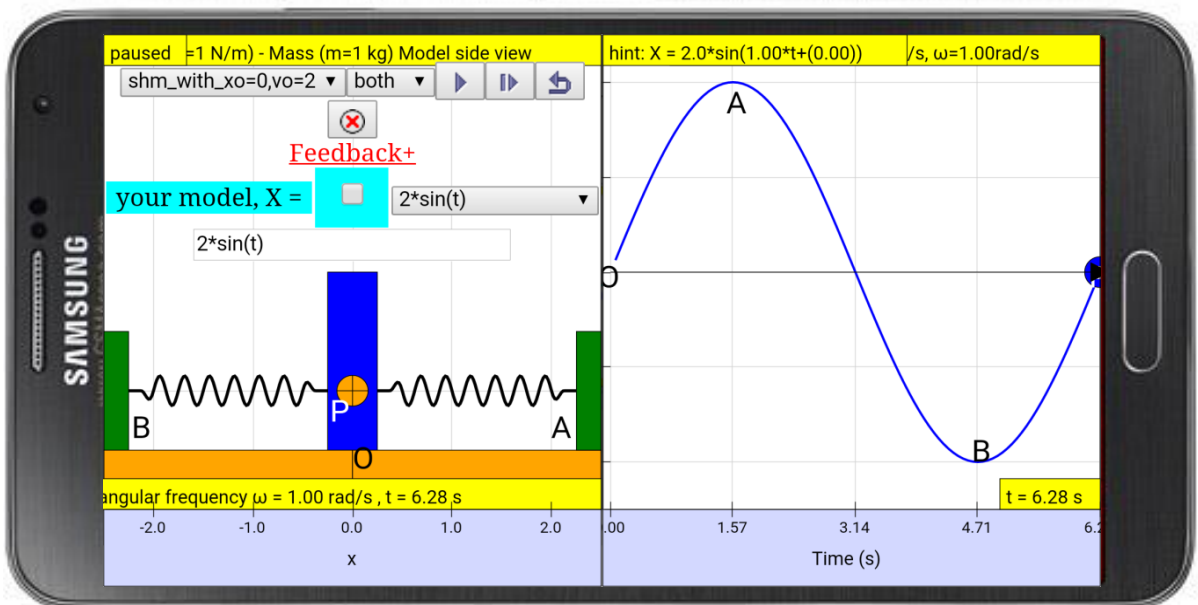

*Figure 2: Same Horizontal Spring Mass Model showing full screen app-like capability, running on the Android Chrome Browser in a Samsung Galaxy Note 3, 5.5 inch screen size diagonally, in the portrait orientation showing with a simple world view layout on the left with only one other scientific view of displacement, x versus time, t graph.*

### 3.2. Teacher design: Gradual buildup of concepts

As teachers using the simulations in the electronic text will want to teach in a more comprehensive manner, connecting earlier concepts to new build-up concepts like kinetic, potential and total energies is needed. Figure 3 shows an oscillating spring-mass system; having the ability to show earlier concepts such as displacement, velocity and accelerations.

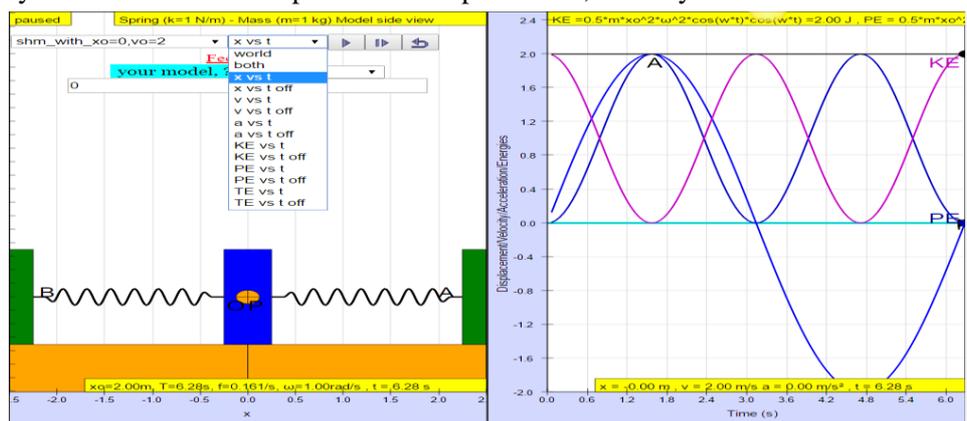

*Figure 3: Horizontal Spring-Mass Model [5] showing a simple world view layout on the left with only one other scientific view of kinetic energy KE, potential energy PE, and total energy TE versus time t, with earlier concepts such as displacement x, versus time, t, velocity v, versus time, t and acceleration a, versus time, t graph. Note that only KE and TE are selected with x versus t.*

---

5   http://iwant2study.org/ospsg/index.php/interactive-resources/physics/02-newtonian-mechanics/09-oscillations/80-shm17

### 3.3. Pedagogical design: progressive mathematical modeling

In view of the phenomenal success of Tracker in allowing students and teachers to represent their understanding of physics through video modeling, I implemented the same kinematic mathematical modeling capability in EjsS models. This was a new approach in teaching and learning with simulations, not implemented in other EjsS and non-EjsS simulations found on the internet. This newly developed progressive mathematical modelling approach in EjsS allows students to propose an initial model where the closeness of fit between simulated data and proposed model suggests an understanding of theoretical application. That is to say, the student may suggest X = 2*cos(t) as an initial model and the simulations immediately draw the plot versus time as a prediction (Radovanović & Sliško, 2013). Clicking the play button then allows the simulations to run as designed for observation. Finally, the apparent mismatch or closeness of fit between model and simulated data can facilitate explanation and discussions. Finally, a "show me" dropdown menu option that displays X = 2*sin(1.00*t+(0.00)) as a generalized solution, is selectable from the drop-down menu, that I believe may help teachers carry out this newly developed progressive mathematical modeling approach in EjsS.

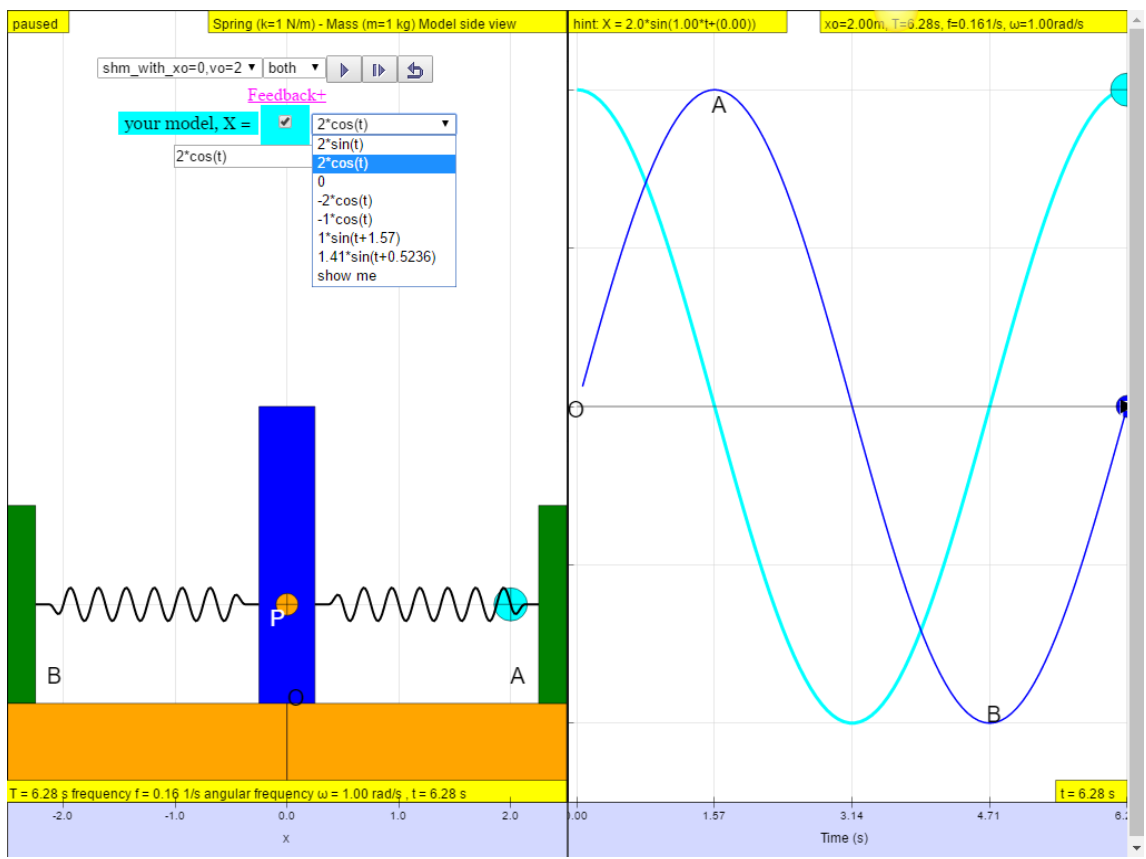

*Figure 4: Horizontal Spring Mass Model [6] with a simple world view layout on the left with only one other scientific view of displacement, x versus time, t graph. Note that student proposed model is selectable via a drop-down menu with some supporting mathematical syntax and a "show me" which determines the generalised form for the motion.*

---

[6] http://iwant2study.org/ospsg/index.php/interactive-resources/physics/02-newtonian-mechanics/09-oscillations/69-shm04

In the chapter on gravitational acceleration and potential[7], similar designs are implemented where students can propose their own model in —for example g = -6.67*500.00/r^2 and ϕ = -6.67*500.00/abs(r) as the gravitational field strength and potential. This can then be plotted in teal alongside the theoretical formula for a gravitational mass M = 500 kg and a gravitational constant G of x10^-11 Nm$^2$/kg$^2$. Again our initial research findings with fifteen grade 11 students in a mainstream junior college setting suggests that this approach can show the meaning of the Red mass M= 500 kg. The fact that this mass creates the gravitational field and potential was previously not clearly understood when explained with paper representations in the lecture notes.

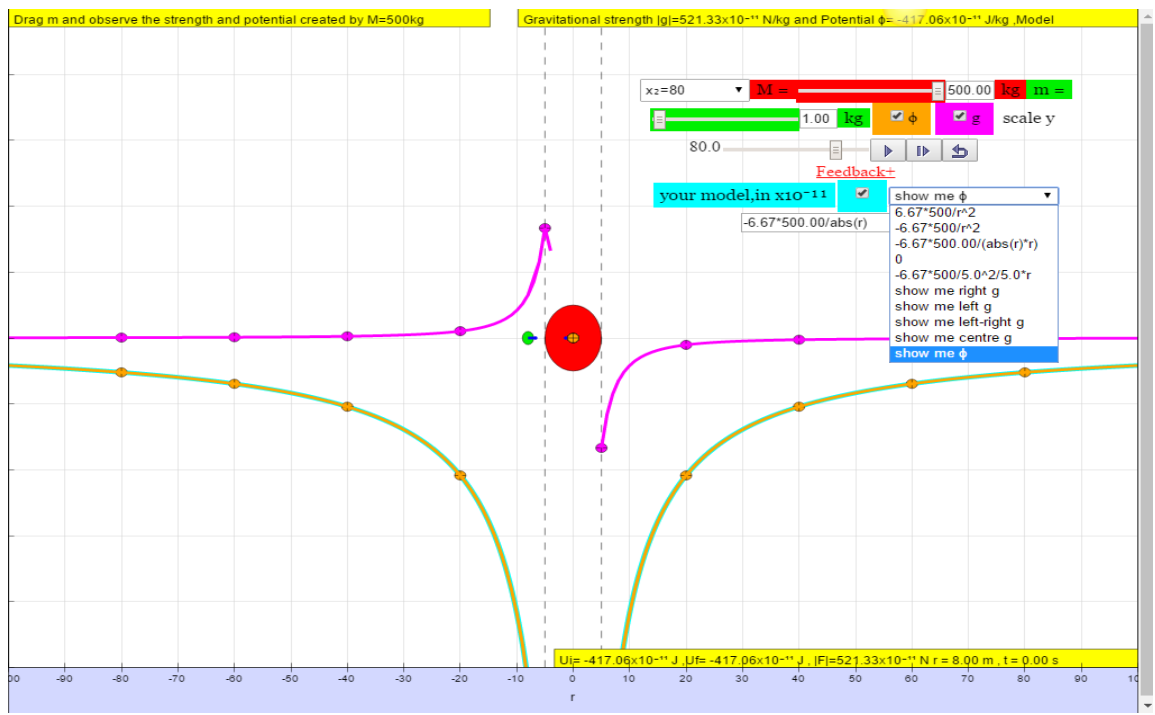

*Figure 5: Gravitational Acceleration and Potential Model [8] with a RED mass M=500 kg creating a gravitational field with a test GREEN mass, m=1 kg, where the field strength g and potential ϕ are simulated. Students then proposed their own initial TEAL model to test if their model matches the theoretical formulae using g = -6.67*500.00/r^2 or ϕ = -6.67*500.00/abs(r) from the drop down menu.*

### 4.  Conclusion

With just these three computer tools (Ejss Modeling Tool, BlueGriffon xhtml editor and Mathtowebonline MathML equation editor) it is possible to create epub3 format electronic books-chapters for computers and tablets that can be published as Apple iBook (Gravity[9], Oscillator[10]), Android Play Book (Gravity[11], Oscillator[12]), and Kindle Book.

---

[7] http://iwant2study.org/ospsg/index.php/interactive-resources/physics/02-newtonian-mechanics/08-gravity/58-gravity06

[8] http://iwant2study.org/ospsg/index.php/interactive-resources/physics/02-newtonian-mechanics/09-oscillations/69-shm04

[9] https://itunes.apple.com/us/book/gravity-advanced-level-gce/id1001442379?mt=11

[10] https://itunes.apple.com/us/book/simple-harmonic-motion/id967139041?mt=11

[11] https://play.google.com/store/books/details/Loo_Kang_Lawrence_Wee_Gravity_Advanced_Level_Physi?id=LS3_CQAAQBAJ

[12] https://play.google.com/store/books/details/Loo_Kang_Lawrence_Wee_Simple_Harmonic_Motion?id=lqGiBgAAQBAJ

Three simulation design principles for electronic book-chapters were discussed above: *3.1 user design: simple & optimum view of screen*, *3.2 teaching design: gradual buildup of concepts and lastly*, and finally, the most significant point, *3.3 pedagogical designs: progressive mathematical modeling*. I hope to have inspired more educators to create interactive and open educational resources for the benefit of all.

## 5. Acknowledgement

I wish express my sincere gratitude for the tireless contributions of Francisco Esquembre, Fu-Kwun Hwang, Wolfgang Christian, Anne Cox, Andrew Duffy, Todd Timberlake and many more in the Open Source Physics community. I have designed much of the above based on their ideas and insights, and I thank the OSP community for which Singapore[13] was honored with 2015-6 UNESCO King Hamad Bin Isa Al-Khalifa Prize for the Use of ICTs in Education. This research is made possible by the eduLab project NRF2015-EDU001-EL021[14], awarded by the Prime Minister Office, National Research Foundation (NRF), Singapore in collaboration with National Institute of Education (NIE), Singapore and the Ministry of Education (MOE), Singapore.

---

[13] http://www.unesco.org/new/en/media-services/single-view/news/the_2015_unesco_king_hamad_bin_isa_al_khalifa_prize_for_the_use_of_icts_in_education_rewards_projects_from_costa_rica_and_singapore/

[14] http://edulab.moe.edu.sg/edulab-programmes/existing-projects/nrf2015-edu001-el021